\documentclass[11pt]{article}
\usepackage{latexsym, amsthm}
\usepackage{amssymb}
\usepackage{amsmath}
\usepackage{graphicx}
\newtheorem{theorem}{Theorem}[section]

\newtheorem{lem}[theorem]{Lemma}

\theoremstyle{definition}
\newtheorem{defn}[theorem]{Definition}

\numberwithin{equation}{section}

\begin{document}
\title{A Generalization of the Poincar\'{e}-Cartan Integral Invariant for a Nonlinear Nonholonomic Dynamical System}
\date{}
\author{Naseer Ahmed\footnote{Mathematics Department, Quaid-I-Azam University, Islamabad, Pakistan}\ ,
\qquad Muhammad Usman \footnote{Mathematics Department, University of Dayton, Dayton, Oh, USA (corresponding author)}}

\date{}
%\email{usmanm@email.uc.edu}
%\keywords{Poincar\'{e}-Cartan Integral
%Invariant, nonlinear nonholonomic dynamical system}
\maketitle
\begin{abstract}
 Based on the d'Alembert-Lagrange-Poincar\'{e} variational
principle, we formulate general equations of motion for mechanical
systems subject to nonlinear nonholonomic constraints, that do not
involve Lagrangian undetermined multipliers. We write these
equations in a canonical form called the Poincar\'{e}-Hamilton
equations, and study a version of corresponding Poincar\'{e}-Cartan
integral invariant which are derived by means of a type of
asynchronous variation of the Poincar\'{e} variables of the problem
that involve the variation of the time. As a consequence, it is
shown that the invariance of a certain line integral under the
motion of a mechanical system of the type considered characterizes
the Poincar\'{e}-Hamilton equations as underlying equations of the
motion. As a special case, an invariant analogous to Poincar\'{e}
linear integral invariant is obtained.
\end{abstract}
\renewcommand{\baselinestretch}{1.7}
\section{Introduction}       % Enter section title between curly braces
The theory of integral invariants (relative or absolute) introduced
by Poincar\'{e} in his pioneering work \cite{Poincare} and extended
by Cartan\cite{Cartan} has found many applications in celestial
mechanics and various other branches of science including analytical
mechanics
\cite{Gomes79,Blacall41,Djukic76,Dobronravov,Gantmacher,Lee47,Feng91,Pars,
Shao93,Whittaker,Yarosheuk}, mathematical physics\cite{Donder},
contact systems \cite{Shu}, fluid mechanics \cite{Ting}, optics and
lasers \cite{Savchenko} etc. In particular, the relative integral
invariants are important for oscillatory systems.
\\It is known \cite{Pars} that the motion of many mechanical systems is restricted by nonholonomic
 constraints so that every position in space is not accessible
and in general,  the variation of the velocity and velocity of
variation
 are not identical. As a consequence, there are two
viewpoints concerning the commutativity of the operation $d$ (actual
variation) and $\delta$ (virtual variation) of kinematical
quantities. The first viewpoint is due to H\"older, Volterra and
Hamel \cite{Neimark-Fufaev} which presumes that $d-\delta$
operation's commutitivity is universally applicable for all
velocities whereas the second viewpoint due to
 Amaldi, Appell, Levi-Civita and Suslov
\cite{GN-HP-NH-94,Neimark-Fufaev} assumes they commute only for
independent generalized velocities.\\So far as the author knows, in
all investigations concerning nonholonomic dynamical systems, the
researchers usually use the first viewpoint in conjunction with the
method of Lagrange undetermined multipliers and employ the
Lagrangian variables or
quasi-coordinates \cite{Djukic76} to study the integral invariant for such systems.\\
In \cite{Arnold}, the author has studied the Poincar\'{e}-Cartan
integral invariant for linear nonholonomic systems in terms of
Lagrangian coordinates and in \cite{Naseer94} author has employed
the Poincar\'{e} formalism based on the Lie theory of continuous
groups of transformations to generalize these results for holonomic
systems. The aim of the present investigation is twofold: (i)to
extend the Poincar\'{e} formalism to include the generalized
variation involving the variation of the time and (ii) to examine
the theory, without appealing to the method of Lagrange undetermined
multipliers, from the second viewpoint \cite{GN-HP-NH-94} for
nonlinear nonholonomic dynamical systems. To achieve this we furnish
a precise definition of the virtual displacement in terms of the
newly introduced parameters \cite{Naseer94} corresponding to the
asynchronous variation and obtain a generalization of the
Poincar\'{e}-Cartan integral invariant for such a systems.
\section{Generalized Variation}       % Enter section title between curly braces
Consider the motion of a holonomic dynamical system whose
configuration at any time $t$ along the actual (real) trajectory is
determined by the set of $C^2$ functions
\begin{equation} \label{eq:xs}
x_{p}(t)\qquad p  =1,2, \cdots,n \end{equation}
 of the time $t$ of
at least class $C^2$ and define the actual displacement in time
$`dt$'  by the relation
\begin{equation}
dx_{p}=\dot{x}_p dt \qquad p  =1,2,\cdots,n \end{equation} Any one
of $\infty^{n}$ configurations infinitely close to the actual
trajectory and compatible with the constraints along a neighboring
curve is defined by
\begin{equation}
x^{*}_{p}(t)=x_{p}(t)+\delta{x}_p  \end{equation}\\
The quantities $\delta{x}_p$, arbitrary functions of the time $t$ of
class $C^1$ , are the variations of the variables $x_{p}(t)$ that
are obtained by keeping the time $t$ fixed. The difference
\begin{equation}
x^{*}_{p}(t)-x_{p}(t)=\delta{x}_p
\end{equation}
is called the synchronous (or simultaneous) virtual variation of the
variable $x_{p}(t)$ and is designated as $\delta$-variation. \\
Turning to a more general variation process, we define the position
of the system in the actual (real) motion by the variables
$x_{p}(t)$ and determine in a varied motion an infinitely closed
position, by the function $x^{*}_{p}(t+\Delta{t})$, where
$\Delta{t}$ is an infinitesimal change in time and is a differential
function of the time t.\newline Taking into account only small
quantities of first order, we get
\begin{equation}
x^{*}(t+\Delta{t})=x^{*}_{p}(t)+\dot{x}^{*}_{p}
\Delta{t}=x_{p}(t)+\delta{x}_p +\dot{x}_{p} \Delta{t},
\end{equation}
where the dot over the letter denotes differentiation with respect
to time $t$. Introducing the symbol $\Delta{x}_p$, we write
\begin{equation}
x^{*}(t+\Delta{t})-x_{p}(t)=\Delta{x}_p =\delta{x}_p+\dot{x}_{p}
\Delta{t},
\end{equation}
which serves to define the variation $\Delta{x}_p$ of the variable
$x_{p}$ where the time varies and the position along the varied path
is not simultaneous to the actual path. These variations are called
asynchronous (or non-simultaneous) virtual variations of the
variable $x_{p}$ and are indicated by the symbol $\Delta$-variation.
Further, the $\Delta$-variation of an arbitrary function
$F(x_{p},t)$ of the class $C^{2}$ in the domain of the variables
$x_{p}$ and time $t$ can be readily obtained as
\begin{equation}
\Delta{F}=\delta{F}+\dot{F} \Delta{t}.
\end{equation}
Since $ \Delta{t}$ is an infinitely small and differentiable
function of the time $t$, by virtue of the relations (2.2) to (2.5),
it can be easily shown that \\
\begin{equation}
d(\delta {x}_p)/dt=\delta\dot{x}_p\ ;\
(d(\Delta{x}_p)/dt=\Delta\dot{x}_p+\ddot{x}_p\Delta{t}),
\end{equation}
which leads to the fact that the velocity of the variation and the
variation of the velocity are the same (not the same)\ according to
the process of $\delta$-variation \ ($\Delta$-variation),
respectively. It is worth mentioning that the relation (2.7) play a
vital role when the investigations are carried out on the basis of
variational principles and their interpretation varies according to
the dynamical system. It is to be noted that the relations (2.8), in
fact, represent the commutative (non-commutative) behavior of the
operation $d/dt$ and $\delta(d/dt\ and\  \Delta)$, respectively.
\\We assume that the relation $d(\Delta t)=\Delta(dt)$ always holds
for the independent variable $t$ whereas it can be shown that
$\Delta \delta{x}_p=\delta\Delta{x}_p$ holds for the dependent
variables ${x}_p$.
\\In order to extend the existing theory of integral invariants we
transform the preceding analysis to the Poincar\'{e} formalism that
is based on the theory of continuous groups of
transformations\cite{Poincare1901}.
\\ Let the group variables ${x}_p$ be holonomically connected to the
cartesian coordinate $u$'s by the relations
\begin{equation}
u_{p}=u_{p}(x_{p},t)\qquad   (p,q=1,2,...,n)
\end{equation}
and the Poincar\'{e} parameters $\eta_{p}(\omega_{p})$ of the actual
(virtual) displacement by the relations\cite{Chetaev41}
\begin{equation}
\eta_{p}=A_{pq}\dot{x}_q+A_p\qquad (\omega_{p}=A_{pq}\delta{x}_q)
\end{equation}
\\where ${A}_{pq}$, ${A}_p$ are functions of the $x's$ and the time
$t$ and the matrix ${A}_{pq}$ is non-singular. In this paper the
following conventions will be observed: (1) the summation convention
is employed throughout the work;(2) the indices will have the range
of the values $\lambda,\mu,\nu=0,1,2,...,n$;\ $p,q,r,s$=1,2,...,.n;\
$i,j,k,l$=1,2,...,$m<n$;\
$\alpha,\beta,\gamma,\delta=(m+1),...,n$.\\We introduce the
following:
 \begin{defn}The change $dG$ of an arbitrary function
 $G(x_{p},t)$ , during the actual displacement $dx_{p}$ in the time
 $dt$ of the system, is determined by the relations \cite{Naseer94}
 \begin{equation}
 dG=[X_{0}G+\eta_{p}X_{p}G]dt\qquad(p=1,2,...,n)
 \end{equation}

where $X_{0}$ and $X_{p}$, characterizing the
 infinitesimal displacement, are the operators defined by
 \begin{equation}
 X_{0}=\frac{\partial}{\partial t}+\xi_{0}^{q}(x_{p})\frac{\partial}{\partial
 x_{q}},\quad X_{p}=\xi_{p}^{q}(x_{r})\frac{\partial}{\partial
 x_{q}}
 \end{equation}
 which form a transitive group of operators if we require that the
 commutators
 \begin{equation}
 (X_{0},X_{p})=X_{0}X_{p}-X_{p}X_{0},\quad (X_{p},X_{q})=X_{p}X_{q}-X_{q}X_{p},\notag
 \end{equation} satisfy the relations
\begin{equation}
 (X_{0},X_{p})=C_{0p}^{q}X_{q}, \quad
 (X_{p},X_{q})=C_{pq}^{r}X_{r}\qquad (p,q,r=1,2,...,n)
 \end{equation}
 Here $C_{0p}^{q}$, $C_{pq}^{r}$, which depend on the $x$'s and the
 time $t$, are the structure constants corresponding to the
 operators $X_{0}$,$X_{p}$.
\end{defn}
 We now give the following:
 \begin{defn}In a simultaneous virtual displacement
 $\delta x_{1},\delta x_{2},...,\delta x_{n}$ of the system, the
 change $\delta G$ in an arbitrary function $G(x_{p},t)$ is
 determined by the formula \cite{Chetaev41}:
 \begin{equation}
\delta G=\omega_{p}X_{p}G.\qquad (p=1,2,...,n)
\end{equation}
Here the quantities $\omega_{p}$ are the parameters corresponding to
the synchronous variation $\delta G$ of the function and are called
the Poincar\'{e} synchronous virtual displacement parameters or
simply the virtual displacement parameters.\\In order to include the
asynchronous variation $\Delta G$ of the function $G(x_{p},t)$
during an infinitesimal time $\Delta t$, we substitute from (2.11)
and (2.14) into the formula (2.7) to obtain
\begin{equation}
\Delta G=(\Delta t)X_{0}G+(\omega_{p}+\eta_{p}\Delta t)X_{p}G \notag
\end{equation}
If the virtual displacements are characterized by the asynchronous
variations $\Delta x_{p}$, then we need to extend definition 2.2.
For this purpose,we use the notation $x_{0}=t,\dot{\eta}_0=1$ and we
write the expression (2.12) for the infinitesimal displacement
operators in the compact form:
\begin{equation}
 X_{\mu}=\xi_{\mu}^\nu(x_{1},...,x_{n})\frac{\partial}{\partial x_{\nu}};\quad
\xi_{0}^{0}=1,\quad \xi_{p}^{0}=0\qquad (\mu,\nu=0,1,2,...,n)
\end{equation}
\newline Further, we set
\begin{equation}
 \omega_{0}=0,\quad \Delta x_{0}=\Delta t=\Omega_{0},\quad \dot{x}_{0}=\eta_{0}=1
\end{equation}
and analogous to the Poincar\'{e} parameters $\omega_{p}$
\cite{Chetaev41}, introduce the new parameters corresponding to
asynchronous variations by the relation
\begin{equation}
 \Omega_{\mu}=\omega_{\mu}+\eta_{\mu}\Omega_{0}
\end{equation}
\newline which was first given in \cite{Arnold,Naseer94}.
\end{defn}
 We now present the
following new definition:
\begin{defn}The variation $\Delta G$ of an
arbitrary function $G(x_{p},t)$, during a virtual displacement
$\Delta x_{1},\Delta x_{2},...,\Delta x_{n}$ asynchronous to the
time $\Delta t=\Omega_{0}$, is described by the formula
\begin{equation}
 \Delta G=\Omega_{\mu} X_{\mu}G \qquad (\mu=0,1,...,n)
\end{equation}
where $\Omega_{0},\Omega_{1},...,\Omega_{n}$, assumed to be
functions of class $C^{2}$, are the parameters of the virtual
displacement corresponding to the asynchronous variation. For future
reference we shall call them  the asynchronous virtual displacement
parameters.

Moreover, in view of (2.16-17), the formula (2.11) can be put in a
more compact form as
\begin{equation}
 dG=[\eta_{\mu}X_{\mu}G]dt
\end{equation}

Since the system is holonomic, by virtue of relations (2.11-13) and
the rule $\delta d=d\delta$, the synchronous variations
$\delta\eta_{p}$ of the parameters of real displacement are given by
\cite{Chetaev41}
\begin{equation}
\delta\eta_{p}=
\frac{d\omega_{p}}{dt}+C_{0q}^{p}\omega_{q}+C_{qr}^{p}\eta_{q}\omega_{r}
\end{equation}
\newline
and, by means of the formula (2.7), it follows that
\begin{equation}
 \Delta\eta_{p}=\delta \eta_{p}+ \dot{\eta}_{p}\Omega_{0}
\end{equation}
\newline
Thus , using relations (2.17) and (2.20), it can readily be shown
that
\begin{equation}
\Delta\eta_{p}=
\dot{\Omega}_{p}-\eta_{p}\dot{\Omega}_{0}+C_{qr}^{p}\eta_{q}\Omega_{r}+C_{0q}^{p}\Omega_{q}
\end{equation}
which represents the asynchronous variations $\Delta\eta_{p}$ of the
parameters of real displacement in terms of the new parameters
$\Omega$'s of possible displacements.
\end{defn}
It is known \cite{NaseerThesis} that the process of
$\delta$-variation and integration commute with each other for
holonomic dynamical systems, but from the preceding analysis it
follows that the non-commutativity of the operations $`\Delta$' and
differentiation '$d$' implies the non-commutativity of the operation
$\Delta$ and integration. That is,
\begin{equation}
\Delta\int_{t_{1}}^{t_{2}}() dt\neq \int_{t_{1}}^{t_{2}} \Delta() dt
\end{equation}
in general, even if the system is holonomic. Precisely, we state the
following:
\begin{lem}
Let $J$ be a functional defined by the integral
\begin{equation}
J=\int_{0}^{t}f dt,
\end{equation}
\\ where $f$ is an arbitrary function of $\eta_{p}, x_{p}$ and
possibly the time $t$. Then the asynchronous variation $\Delta J$ of
the functional $J$ is given by
\begin{equation}
\Delta J=\int_{0}^{t}(\Delta f+f \dot{\Omega}_{0}) dt.
\end{equation}
\end{lem}
For the proof of the lemma, readers are referred to \cite{Naseer94}.
However, one can see that the non-commutativity of the
$\Delta$-operation and the integration is obvious due to the
presence of the quantity $f \dot{\Omega}_{0}$ in the expression
(2.25).

In a special case when the variation is synchronous, we have $\Delta
t=\Omega_{0}=0$. This, together with (2.17) implies that
$\Delta\equiv\delta$ and (2.25) includes the well-known result of
the commutativity of $\delta$-operation and integration for a
holonomic dynamical system. But it is to be noted carefully that for
nonholonomic system these results vary in a different way. See
\cite{GN-HP-NH-94} for a discussion of this case.
\section{The two viewpoints}
From the analysis discussed in the preceding section, it follows
that the commutativity (or non-commutativity) of the operations $d$
with $\delta$ (or $\Delta$) is expressed by the transpositional
relations (2.20) and (2.22) respectively, which play a crucial role
in the dynamics of the nonholonomic systems. In fact, the study of
integral invariants for such systems is intimately connected with
these relations, since they involve certain variational procedure.
In the sequel we briefly discuss  the viewpoints about the validity
of these relations which allow us to obtain a generalization of the
theory of integral invariants.

Let us consider the dynamical system whose configuration is
determined by the group variables $x_{p}$ and moves subject to the
nonholonomic constraints expressed by the $(n-m)$ independent
equations
\begin{equation}
f_{\alpha}(\eta_{p},x_{p},t)=0\qquad (p=1,2,...,n ;
\alpha=m+1,...,n),
\end{equation}
where the functions $f_{\alpha}$ are not necessarily linear in the
$\eta_{p}$'s and that the virtual displacement parameters
$\omega_{p}$'s satisfy the Chetaev's relations \cite{Chetaev41}
\begin{equation}
\frac{\partial f_{\alpha}}{\partial \eta_{p}}\omega_{p}=0.
\end{equation}
\\We further assume that the constraint equations (3.1) are
expressible in the form
\begin{equation}
f_{\alpha}(\eta_{p},x_{p},t)=\eta_{\alpha}-\phi_{\alpha}(\eta_{i},x_{p},t)=0,
\end{equation}
 \begin{center}$(p=1,2,...,n; i=1,2,...,m<n;
\alpha=m+1,...,n)$\end{center} and the $\omega$'s satisfy the
relations \cite{GN-HP-NH-94}
\begin{equation}
\omega_{\alpha}=\frac{\partial \phi_{\alpha}}{\partial
\eta_{i}}\omega_{i}
\end{equation}
where we designate $\eta_{\alpha} (\eta_{i})$ and $\omega_{\alpha}
(\omega_{i})$ as the dependent (independent) Poincar\'{e} parameters
of real and virtual displacements respectively.

In view of (3.2) or (3.4), we observe that the presence of
nonholonomic constraints (3.1) or (3.3) leads to the fact that the
$\omega_{p}$'s are not independent. This reveals that the parameters
of virtual displacement are not determined uniquely. Thus, there
exists some arbitrariness in the determination of the derivatives
$\frac{d\omega_{p}}{dt}$. Consequently, in using relation (2.20), we
can adopt either of the two viewpoints given in the following:

i) According to H\"{o}lder, Volterra and Hamel, (2.20) may be used
for the $\delta$-variation of all the parameters $\eta_{p}$ of real
displacement whether the system is holonomic or not. This, in view
of the formula (2.14), allows us to express the variations $\delta
f_{\alpha}$ of $f_{\alpha}$ for the constraints (3.1) and (3.3) in
the form \cite{GN-HP-NH-94}:
\begin{equation}
\delta f_{\alpha}=\left[X_{p} f_{\alpha}
+(C_{0p}^{r}+C_{qp}^{r}\eta_{q})\frac{\partial f_{\alpha}}{\partial
\eta_{r}}-\frac{d}{dt}\left(\frac{\partial f_{\alpha}}{\partial
\eta_{p}}\right)\right ]\omega_{p}
\end{equation}
and
\begin{equation}
\delta f_{\alpha}= \delta \eta_{\alpha}-\delta
\phi_{\alpha}=A_{i}^{\alpha} \omega_{i}
\end{equation}
respectively, by using the conditions (3.2) and (3.4). Here the
quantities $A_{i}^{\alpha}$ are determined by the relations
\begin{equation}
\begin{split}
A_{i}^{\alpha}&=\frac{d}{dt}\left(\frac{\partial
\phi_{\alpha}}{\partial \eta_{i}}\right ) -X_{i}\phi_{\alpha}-
\frac{\partial \phi_{\beta}}{\partial
\eta_{i}}X_{\beta}\phi_{\alpha}+(C_{0i}^{\alpha}+C_{qi}^{\alpha}\eta_{q})
+(C_{0\beta}^{\alpha}+C_{q\beta}^{\alpha}\eta_{q})\frac{\partial
\phi_{\beta}}{\partial \eta_{i}}\\
&-\frac{\partial \phi_{\alpha}}{\partial \eta_{j}}
\{(C_{0i}^{j}+C_{qi}^{j}\eta_{q})+ (C_{0 \beta}^{j}+C_{q
\beta}^{j}\eta_{q})\frac{\partial \phi_{\beta}}{\partial
\eta_{i}}\}\\
&(q=1,2,...,n ;i,j=1,2,...,m<n;\alpha,\beta=m+1,...,n)
\end{split}
\end{equation}

ii) According to Amaldi, Levi-Civita and Suslov as discussed in
\cite{GN-HP-NH-94}, the relations (2.20) hold for nonholonomic
systems. They may be used only for the variation $\delta \eta_{i}$
of the independent real parameters $\eta_{i}$ together with the
assumption that the $\delta$-variation of the $f_{\alpha}$ vanishes.
Precisely,
\begin{equation}
\delta\eta_{i}=
\frac{d\omega_{i}}{dt}+C_{0q}^{i}\omega_{q}+C_{qr}^{i}\eta_{q}\omega_{r}\quad;\quad
\delta f_{\alpha}=0
\end{equation}
for the constraints of the type (3.3).

The relations $\delta f_{\alpha}=0$ enable us to obtain the
synchronous virtual variations of the dependent $\eta$'s. To
distinguish them from the variation $\delta f(\eta_{p},x_{p},t)$ of
arbitrary function of all the $\eta$'s, $x$'s and $t$, we employ the
symbol $\delta^*$ to denote the variation of $f$ when computed in
terms of the independent Poincar\'{e} parameters by means of (3.3).
Thus, from (2.14) and (3.4), it follows that\\
\begin{equation}
\delta^* f_{\alpha}=\delta^* \eta_\alpha-\delta^*
\phi_\alpha=(A_j^\alpha)^* \omega_j
\end{equation}
Here $(A_j^\alpha)^*$ are given by
\begin{equation}
(A_j^\alpha)^*=\frac{d}{dt}(\frac{\partial
\phi_\alpha}{\partial\eta_j})-X_j^*\phi_\alpha+(K_{0j}^\alpha+K_{ij}^\alpha
\eta_i+K_{\beta j}^{\alpha} \phi_\beta)-(K_{0j}^{k}+K_{ij}^k
\eta_i+K_{\beta j}^{k} \phi_\beta)\frac{\partial
\phi_\alpha}{\partial \eta_k}
\end{equation}
where
\begin{eqnarray}
K_{0j}^{k}&=&C_{0j}^k+C_{0\beta}^k \frac{\partial
\phi_\beta}{\partial\eta_j}\quad;\quad
K_{0j}^{\alpha}=C_{0j}^{\alpha}+C_{0\beta}^{\alpha} \frac{\partial
\phi_\beta}{\partial\eta_j}\\
K_{qj}^{k}&=&C_{qj}^k+C_{q \beta}^k \frac{\partial
\phi_\beta}{\partial\eta_j}\quad;\quad
K_{qj}^{\alpha}=C_{qj}^{\alpha}+C_{q\beta}^{\alpha} \frac{\partial
\phi_\beta}{\partial\eta_j}\\
and \notag\\
X_j^{*}&=&X_j+\frac{\partial \phi_\beta}{\partial\eta_j}X_\beta
\end{eqnarray}
\section{Equations of Motion without Lagrange Multipliers}
We start with the general equation of dynamics expressing the
d'Alembert-Lagrange-Poincar\'{e} principle given by
\cite{GN-HP-NH-94}
\begin{equation}
\left[\frac{d}{dt}\left(\frac{\partial
L}{\partial\eta_p}\right)-C_{0p}^q\frac{\partial L}{\partial\eta_q}
-C_{qp}^r \eta_{q}\frac{\partial L}{\partial\eta_r}-X_{p}L\right
]\omega_p=0
\end{equation}
where $L(\eta_{p},x_{p},t)$ is the Lagrangian function that
describes the dynamical behavior and the $\omega_{p}$ determine the
virtual displacement of the system. We assume that the system moves
subject to the nonlinear nonholonomic constraints (3.3) and adopt
the second viewpoint. Since the $\omega$'s are not independent and
satisfy the conditions (3.4), the general equation of dynamics
(4.1), as it stands, does not yield the equation of motion.
Therefore, we write (4.1)in the form
\begin{equation}
\begin{split}
&\left[\frac{d}{dt}\frac{\partial L}{\partial \eta_{j}}+
\frac{\partial \phi_{\beta}}{\partial
\eta_{j}}\frac{d}{dt}\frac{\partial L}{\partial
\eta_{\beta}}-(C_{0j}^{q}+C_{0\beta}^{q}\frac{\partial
\phi_{\beta}}{\partial \eta_{j}})-(C_{q
j}^{r}+C_{q\beta}^{r}\frac{\partial \phi_{\beta}}{\partial
\eta_{j}})\eta_q\right.\\
&-\left.(X_j L+\frac{\partial \phi_{\beta}}{\partial \eta_j}X_\beta
L)\right]\omega_j=0 \\
&(q=1,2,...,n ;i,j=1,2,...,m<n;\alpha,\beta=m+1,...,n)
\end{split}
\end{equation}
Where we have used the relations (3.4) after separating the sum over
$p$ from 1 to $n$ into the sums over $j$ from 1 to $m$ and over
$\beta$ from $(m+1)$ to $n$.\\
After expressing the $\eta_{\alpha}$'s in terms of $\eta_j$'s we
call the resulting function $L^{*}(\eta_j,x_p,t)$. Thus,
\begin{equation}
L^{*}(\eta_i,x_p,t)=L(\eta_j,\phi_\alpha (x_p,\eta_j,t),x_p,t),
\end{equation}
as a result of operating by $X_p$ and differentiating partially with
respect to $\eta_j$, we have
\begin{equation}
X_p L^{*}=X_p L+\frac{\partial L}{\partial \eta_{\alpha}}X_p
\phi_\alpha \qquad \frac{\partial L}{\partial
\eta_{j}}=\frac{\partial L^{*}}{\partial \eta_{j}}-\frac{\partial
L}{\partial \eta_{\alpha}}\frac{\partial \phi_\alpha}{\partial
\eta_j}
\end{equation}
Substituting from (4.3-4) into (4.2) and performing some
mathematical manipulations the general equation of dynamics takes
the form
\begin{small}
\begin{equation}
\begin{split}
&\left[\frac{d}{dt}\big(\frac{\partial L^{*}}{\partial
\eta_{j}}\big)- \left\{(C_{0j}^k+C_{0\beta}^k\frac{\partial
\phi_\beta}{\partial \eta_j})+(C_{qj}^k+C_{q\beta}^k\frac{\partial
\phi_\beta}{\partial \eta_j})\eta_{q}\right\}\frac{\partial
L^{*}}{\partial \eta_{k}} \right.\\
&-(X_{j}+\frac{\partial \phi_\beta}{\partial \eta_j}X_{\beta})L^{*}
-(\frac{\partial L}{\partial
\eta_{\alpha}})^{*}\left\{\frac{d}{dt}(\frac{\partial
\phi_{\alpha}}{\partial \eta_{j}})-(X_{j}+\frac{\partial
\phi_\beta}{\partial \eta_j}X_{\beta})\phi_{\alpha}\right.\\
&+(C_{0j}^\alpha+C_{0\beta}^\alpha\frac{\partial
\phi_\beta}{\partial \eta_j})
+(C_{qj}^\alpha+C_{q\beta}^\alpha\frac{\partial \phi_\beta}{\partial
\eta_j})\eta_{q}-\frac{\partial \phi_\alpha}{\partial
\eta_k}\left((C_{0j}^k+C_{0\beta}^k\frac{\partial
\phi_\beta}{\partial
\eta_j})\right.\\
&+\left.\left.\left.\left.(C_{qj}^k+C_{q\beta}^k\frac{\partial
\phi_\beta}{\partial
\eta_j})\eta_{q}\right)\right\}\right]\right.\omega_{j}=0
\end{split}
\end{equation}
\end{small}
From the independence of the $\omega$'s and (3.10-12), the last
result leads to
\begin{equation}
\frac{d}{dt}\big(\frac{\partial L^{*}}{\partial \eta_{j}}\big)-
(K_{0j}^k+K_{q j}^k \eta_{q})\frac{\partial L^{*}}{\partial
\eta_k}-X_{j}^{*} L^{*}-(A_{j}^{\alpha})^*\big(\frac{\partial
L}{\partial \eta_{\alpha}}\big)^{*}=0,
\end{equation}
which are the required equations of motion independent of the
Lagrange undetermined multipliers for the nonlinear nonholonomic
dynamical system. These $m$ equations determine the values of
$\eta_{j} (j=1,2,...,m)$ which, by virtue of the constraint
equations (3.3), allow us to determine the values of the remaining
$(n-m)$ $\eta_{\alpha}$'s as functions of the $x_{p}$'s and $t$. The
values of $x_{p}$ are then calculated from the equations
\begin{equation}
\dot{x}_{p}=X_0 x_{p}+\eta_q X_q x_p
\end{equation}
which are obtained from (2.11) by setting $G=x_p$.
\\ \indent In order to obtain the canonical form of equations (4.6),
we introduce the generalized momenta $y_p$ by the relations
\begin{equation}
y_p=\frac{\partial L}{\partial \eta _p}
\end{equation}
and we assume the transformations (4.8) to be invertible so that the
$\eta_p$'s can be expressed in the form
\begin{equation}
\eta_p=\eta_p(x_q,y_q,t) \qquad (p,q=1,2,...,n)
\end{equation}
We note that (4.8) and (4.9) yield, respectively, $y$'s and $\eta$'s
as a linear function of $\eta$'s and $y$'s.\\

Since all the $\eta$'s are not independent, it follows that $y$'s
are not independent as well. Therefore, we cannot work in the
phase-space of $2n$ variables defined in \cite{Naseer94}, that is,
\begin{defn}
A set of $2n$ independent quantities $(x_p,y_p)$ is said to form the
phase space if the conditions
\begin{equation}
\delta t=0,\quad \omega_p\neq0,\quad \delta\eta_p\neq 0,\quad \delta
y_p\neq 0 \qquad (p=1,2,...,n)\notag
\end{equation}
hold throughout the motion of the system.
\end{defn}

To overcome this problem we introduce the new independent variables
$y^*_j$ that are defined by means of equations of transformation
\begin{equation}
y^*_j=\frac{\partial L^{*}}{\partial \eta_{j}}
\end{equation}
which are assumed to be invertible and allow to express the
$\eta_j$'s as functions of $y^*_j, x_p$ and time $t$ in the form
\begin{equation}
\eta_j=\eta_j(x_p,y^*_j,t) \qquad (j=1,2,...,m<n;p=1,2,...,n)
\end{equation}
Note that the constraint equations (3.3) are expressed in terms of
the independent $y^*_j$'s by means of the equations (4.11).

We now give the following:
\begin{defn}
The $2m$ quantities $(x_1,...,x_m;y^{*}_1,...,y^{*}_m)$ are said to
form a $2m$-dimensional reduced phase space provided that they are
independent and satisfy the conditions
\begin{equation}
\delta t=0;\quad \omega_j\neq0;\quad \delta\eta_j\neq 0;\quad \delta
y^{*}_j\neq 0
\end{equation}
throughout the motion of the dynamical system under the nonlinear
nonholonomic constraints (3.3) with (3.4).
\end{defn}

Let us consider the motion of the system in this $2m$-dimensional
reduced phase space and introduce the Hamiltonian function
$H^{*}(x_p,y^{*}_j,t)$ corresponding to Lagrangian $L^*$
\begin{equation}
H^{*}(x_p,y^{*}_j,t)=\eta_jy^{*}_j-L^*(x_p,\eta_j,t)
\end{equation}
Performing the $\delta$-variation according to (2.14), in
conjunction with (4.10), we have
\begin{equation}
\delta H^{*}(x_p,y^{*}_j,t)=\delta[\eta_jy^{*}_j-L^*(x_p,\eta_j,t)]
\end{equation}
On one hand, using the second viewpoint  and (4.10), we have
\begin{equation}
\delta^{*} H^{*}(x_p,y^{*}_j,t)=\eta_j\delta^{*}y^{*}_j-\omega_p X_p
L^* \notag
\end{equation}
Breaking the sum over the index $p$ from $1$ to $n$ into the sums
over the indices $j$ from $1$ to $m$ and $\alpha$ from $(m+1)$ to
$n$ and using conditions (3.4) and (3.13), the last result assumes
the form
\begin{equation}
\delta^{*} H^{*}(x_p,y^{*}_j,t)=\eta_j\delta^{*}y^{*}_j-\omega_j
X^{*}_j L^* \notag
\end{equation}
which, in view of equation (4.6), becomes
\begin{equation}
\delta\big[\eta_jy^{*}_j-L^*(x_p,\eta_j,t)]=\eta_j \delta
y^{*}_j-\omega_j[\dot{y}^{*}_j-(K^{k}_{0j}+K^{k}_{qj}\eta_q)\frac{\partial
L^{*}}{\partial \eta_{k}}-(A^{\alpha}_j)^*(\frac{\partial
L}{\partial \eta_{\alpha}})^{*}\big]
\end{equation}
On the other hand, we can find
\begin{equation}
\delta H^{*}(x_p,y^{*}_j,t)= \frac{\partial H^{*}}{\partial
y^{*}_j}\delta y^{*}_j+\omega_p X_p H^{*} \notag
\end{equation}
which is equivalent to
\begin{equation}
\delta H^{*}(x_p,y^{*}_j,t)= \frac{\partial H^{*}}{\partial
y^{*}_j}\delta y^{*}_j+\omega_j X^{*}_p H^{*}
\end{equation}
where we have used (3.4) and (3.12). \\From (4.15) and (4.16), it
follows that
\begin{equation}
( \frac{\partial H^{*}}{\partial y^{*}_j}-\eta_j)\delta
y^{*}_j+\big[X^{*}_j
H^{*}+\dot{y}^{*}_j-(K^{k}_{0j}+K^{k}_{qj}\eta_q)\frac{\partial
L^{*}}{\partial \eta_k}-(A^{\alpha}_j)^{*}(\frac{\partial
L}{\partial \eta_{\alpha}})^{*}\big]\omega_j=0\notag
\end{equation}
since all the $\delta y^*_j$ and $\omega_j$ are independent, the
coefficient of each of the $\delta y^*_j$ and $\omega_j$ vanishes
and we obtain
\begin{equation}
\begin{split}
\eta_j=\frac{\partial H^*}{\partial y_j^*};\quad
\dot{y}^*_j=-X^*_jH^*+(K^{k}_{0j}+K^{k}_{qj}\eta_q)\frac{\partial
L^{*}}{\partial \eta_k}-(A^{\alpha}_j)^{*}(\frac{\partial
L}{\partial \eta_{\alpha}})^{*}\\
(j,k=1,2,...,m;\alpha=m+1,...,n;q=1,2,...,n)
\end{split}
\end{equation}

The equations (4.17) together with (4.7) are the required
Poincar\'{e}-Hamilton(PH) equations of motion for the nonlinear
nonholonomic dynamical system. They together with the constraint
equations (3.3) determine the $(n+m)$ quantities
$x_1,x_2,...,x_n;y_1,y_2,...,y_m$. In fact, we can find
$\eta_j(j=1,2,...,m)$ as functions of $x_p,\ y_j\ \text{and}\ t$ and
then substituting the values of $\eta_j$ into the equations (3.3),
determine the remaining $\eta_\alpha(\alpha=m+1,...,n)$. In this way
all the $\eta_j$'s are determined and then, by using (4.7), all the
$x_p$'s can be obtained as functions of the time $t$.
\section{Poincar\'{e}-Cartan Integral Invariant of Nonholonomic Dynamical System}
We now turn to the investigation of the Poincar\'{e}-Cartan integral
invariants for the conservative nonlinear nonholonomic system whose
motion is determined by the Poincar\'{e}-Hamilton (PH) equations
(4.17) with (4.7). In order to achieve our goal we need to compute,
subject to the constraints (3.3) with (3.4), the asynchronous
variation of the action integral defined by
\begin{equation}
S=\int_{t_1}^{t_2} L dt
\end{equation}
where $Ldt$ expresses the small element of action with $L$,
describing the dynamical behavior of the system, as the Lagrangian
function of all the $x$'s, $\eta$'s and possibly the time $t$.

Let us now consider the motion of the system along the real (actual)
trajectory between the two positions $P_1$ and $P_2$ corresponding
to the initial and final instants $t_1$ and $t_2$, respectively. We
assume that the varied path is determined by means of
$\Delta$-variation in which not only the coordinates but also the
time varies at the initial and final moments of the motion. Since
the Lagrangian is a function of class $C^2$ with respect to all of
its arguments, using Lemma 2.1, we perform the $\Delta$-variation of
(4.1) according to (2.25) to get
\begin{equation}
\Delta S=\int_{t_1}^{t_2} (\Delta L+L\dot{\Omega}_0) dt \notag
\end{equation}
which, by virtue of (A.4) from the APPENDIX, assumes the form
\begin{eqnarray}
\Delta S&=&\big(\frac{\partial L^*}{\partial
\eta_j}\omega_j\big)|_{t_1}^{t_2}+\int_{t_1}^{t_2}
\bigg(\frac{\partial L}{\partial
{\eta}_p}\dot{\eta}_p\Omega_0+\eta_\mu \Omega_0X_\mu\notag
L-\Omega_0\dot{L}\bigg)dt\\
 &+&\int_{t_1}^{t_2}(L\Omega_0)^{.}
dt -\int_{t_1}^{t_2}\left[\frac{d}{dt}(\frac{\partial
L^*}{\partial\eta_j})-(K_{0j}^k+K_{qj}^k \eta_q )\frac{\partial
L^*}{\partial \eta_k}-X^*_j L^* \right.\notag\\
&-&\left.(A^{\alpha}_j)^*(\frac{\partial L}{\partial
\eta_{\alpha}})^*\right]\omega_j dt
\end{eqnarray}
where the superscript $"*"$ shows that the quantities are expressed
in terms of the independent Poincar\'{e} parameters of real
displacement.

Since the Poincar\'{e} equations (4.6) hold along the real(actual)
trajectory, using (2.11), (2.16) and (4.3), the last integral on the
right hand side of (5.2) vanishes, and we are left with
\begin{eqnarray}
\Delta S=\left.\left(\frac{\partial L^*}{\partial
\eta_j}\omega_j+L^* \Omega_0\right)
\right|^{t_2}_{t_1}+\int_{t_1}^{t_2}(\frac{\partial
L}{\partial\eta_p} \dot{\eta_p} \Omega_0+\eta_{\mu} \Omega_0 X_\mu
L&-&\frac{\partial L}{\partial \eta_p}\dot{\eta_p}\Omega_0\notag\\
&-&\eta_\mu \Omega_0 X_\mu L)dt\notag
\end{eqnarray}
And, in view of (4.10), simplifies to
\begin{equation}
\left.\Delta S=\left(y^*_j \omega_j+L^*
\Omega_0\right)\right|^{t_2}_{t_1}
\end{equation}
Using the relations (2.17) and (4.13), the last result becomes
\begin{equation}
\left.\Delta S=\left(y^*_j \Omega_j+H^*
\Omega_0\right)\right|^{t_2}_{t_1}
\end{equation}
where the Hamiltonian function $H^*$ is computed along the actual
trajectory of the system. Denoting it by $-y^*_0$, the asynchronous
variation $\Delta S$ of $S$ can be put in a more concise form
\begin{equation}
\left.\Delta
S=y^*_{\lambda}\Omega_{\lambda}\right|^{t_2}_{t_1}\qquad
(\lambda=0,1,2,...,m)
\end{equation}

If we take $G=x_{\lambda}$ in formula (2.18), we find that
\begin{equation}
\Delta x_{\lambda}=\Omega_{\mu} X_{\mu} x_{\lambda}\qquad (\lambda,
\mu=0,1,2,...,m),
\end{equation}
and , by using (2.12) it can readily be shown that $\Omega_{\mu}$
are linear combinations of the quantities $\Delta x_{\lambda}$ in
the form
\begin{equation}
\Omega_{\mu}=\zeta^{\mu}_{\lambda}\Delta x_{\lambda}
\end{equation}
where $\left \|\zeta^{\lambda}_{\mu} (x_{\nu})\right \|$ is the
inverse of the matrix $\left\|\xi^{\mu}_{\lambda}(x_{\nu})\right\|$.
Then the variational equation (5.5) becomes
\begin{equation}
\left.\Delta S=(y^*_{\lambda} \xi^{\lambda}_{\mu} \Delta
x_{\mu})\right|^{t_2}_{t_1}=\left.(y^*_{\mu}\Delta
x_{\mu})\right|^{t_2}_{t_1}
\end{equation}
Equation (5.8) yields in concise form the variation of the
functional $S$ in the space of $(2m+1)$ variables
$(x_1,x_2,...,x_m;y^*_1,y^*_2,...,y^*_m;t)$ called the "Extended
Reduced Phase-Space" and may be applied to any functional of this
type.

Let us now consider the motion of the system in this space. We note
the fact that corresponding to the different initial values of the
$x$'s and $y$'s we obtain a set of initial points at the initial
time $t_1$ and through each point we can draw the appropriate real
paths satisfying (4.17) and giving rise to a set of terminal points
at the other end of each real path at the same terminal time $t_2$.
If the initial set of points forms a closed curve $C_1$,
correspondingly, the terminal set of points will also form a closed
curve $C_2$ giving a tube of real trajectories.

Thus integration of (5.5) along a closed curve $C$ (that is, the
locus at any time $t$ of the points which are initially on $C_1$)
that passes from $C_1$ to $C_2$, yields
\[
0=\oint_C \Delta S=\oint_C(\left.y^*_{\lambda}
\Omega_{\lambda}\right|^{t_2}_{t_1})=\oint_C
({y^*_{\lambda}}^{(2)}{\Omega_{\lambda}}^{(2)}-{y^*_{\lambda}}^{(1)}{\Omega_{\lambda}}^{(1)}),
\]
where
${y^*_{\lambda}}^{(1)},{y^*_{\lambda}}^{(2)},{\Omega_{\lambda}}^{(1)},{\Omega_{\lambda}}^{(2)}$
are the values of the $y^*$'s and the $\Omega$'s at the time $t_1$
and $t_2$ respectively. In view of the continuity of the motion, the
last result may be written as
\[
\oint_{C_1} y^*_{\lambda} \Omega_{\lambda}=\oint_{C_2} y^*_{\lambda}
\Omega_{\lambda},
\]
which leads to
\[
I=\oint_C (y^*_j \Omega_j-H^* \Omega_0),
\]
taken along a closed contour $C$ remain invariant during an
arbitrary displacement with deformation of the system. Thus, we have
the following:

\begin{theorem} The line integral
\begin{equation}
 I=\oint_C (y^*_j \Omega_j-H^* \Omega_0),
\end{equation}
along an arbitrary closed curve $C$ remains invariant with arbitrary
deformation of this curve along the tube of real trajectories of a
conservative nonlinear nonholonomic dynamical system whose motion is
governed by the equations (4.17) provided that the relations (3.8)
hold.
\end{theorem}
We remark that the integral (5.9) is a generalized form of the
relative integral invariant of Poincar\'{e}-Cartan given in
\cite{Naseerpreprint}. It is shown
\cite{Naseer94,Naseerpreprint,Arnold} that these integrals are
important in the study of analytical dynamics of holonomic system,
since their invariance yields their relationship with the
Poincar\'{e}-Hamiltonian system. In what follows, we shall show that
this relationship can also be established for nonlinear nonholonomic
dynamical systems as well.

\section{Poincar\'{e}-Hamiltonian Systems and the Poincar\'{e}-Cartan Integral Invariant.}
In the preceding section we have established the integral invariant
of a nonlinear nonholonomic dynamical system. Now we shall discuss
the consequences of the invariant property of the integral (5.9).
Precisely, we prove the converse of Theorem 5.1 by making use of the
property of asynchronous variation as discussed in section 2.

We start with the assumption that the generalized
Poincar\'{e}-Cartan integral (5.9) is invariant with respect to the
tube of real trajectories of the system whose motion is governed by
the set of equations of the form
\begin{equation}
\eta_j=\psi_j(x_p,y^*_k,t)
\end{equation}
\begin{equation}
\dot{y}^*_j=\phi_j(x_p,y^*_k,t)\quad (j,k=1,2,...,m;p=1,2,...,n)
\end{equation}
where $\psi_j$ and $\phi_j$ are arbitrary functions to be
determined.

The invariance of the integral (5.9) along the tube of real
trajectories, in accordance with (2.19), (6.1) and (6.2) implies
that
\[
dI=0
\]
Thus, we have
\begin{equation}
0=d\oint (y^*_j \Omega_j-H^* \Omega_0)=\oint_C (dy^*_j
\Omega_j+y_j^* d\Omega_j-dH^*\Omega_0-H^* d\Omega_0)
\end{equation}
Recalling that  relation $d(\Delta t)=\Delta(dt)$ it follows that
\[
H^* d\Omega_0=H^* d(\Delta t)=H^* \Delta(dt)=\Delta (H^* dt)-\Delta
H^* dt
\]
Combining this with (2.22), and the identity
\[
y^*_j \eta_j d\Omega_0=y^*_j \eta_j d(\Delta t)=y^*_j \eta_j
\Delta(dt)=\Delta(y^*_j \eta_j dt)-\Delta(y^*_j \eta_j)dt,
\]
we obtain
\[
\begin{split}
0=d\oint \left[dy^*_j \Omega_j+y^*_j \eta_j
d\Omega_0+\left(y^*_j\Delta \eta_j-y^*_j C^j_{qr} \eta_q
\Omega_r-y^*_j C^j_{0q}\Omega_q\right)dt\right.\\
\left.-dH^* \Omega_0-\Delta (H^* dt)+\Delta H^* dt \right]
\end{split}
\]
which takes the form
\[
\begin{split}
0=d\oint \Delta \left[(y^*_j \eta_j-H^*)dt\right]+ \oint \left[
dy^*_j \Omega_j+y^*_j\Delta \eta_jdt -\Delta(y^*_j\eta_j)dt\right.\\
\left.-y^*_j C^j_{qr} \eta_q \Omega_rdt-y^*_j C^j_{0q}\Omega_q
dt-dH^* \Omega_0+\Delta H^* dt\right]
\end{split}
\]
Since the motion is represented by the closed curve $C$ which is
completely arbitrary, the first integral on the right hand side
 vanishes. Thus we have
\begin{eqnarray}
0=d\oint \left[dy^*_j \Omega_j+y^*_j \Delta \eta_j dt-\Delta(y^*_j
\eta_j)dt -y^*_j\eta_q  C^j_{qr} \Omega_r dt-y^*_j C^j_{0q}\Omega_q
dt\right.\notag\\
-\left.dH^* \Omega_0+\Delta H^*dt\right]\notag
\end{eqnarray}
Interchanging the indices $r$ and $q$ and separating the sum over
the index $q=1,2,...,n$ into the sums over the indices $i=1,...,m$
and $\alpha=m+1,...,n$, we have
\[
\begin{split}
0&=d\oint \left[dy^*_j \Omega_j+y^*_j \Delta \eta_j dt-\Delta(y^*_j
\eta_j)dt -y^*_j\left(C^j_{ri}\eta_r+C^j_{0i}\right)\Omega_i
dt\right.\\
&\left.-y^*_j\left(C^j_{r\alpha}\eta_r+C^j_{0\alpha}\right)\Omega_{\alpha}
dt-dH^*\Omega_0+\Delta H^* dt\right]
\end{split}
\]
Using the formulae(2.18-19) and performing some algebra we get
\begin{small}
\begin{eqnarray*}
0=\oint \left[\left(\dot{y}^*_i-C_{0i}^jy^*_j-C^j_{ri}\eta_r
y^*_j\right)\Omega_idt+y^*_j\Delta\eta_jdt-\Delta y^*_j \eta_jdt
-y^*_j
\Delta \eta_j dt\right.\\
-y^*_j\left(C^j_{0\alpha}+C^j_{r\alpha}\eta_r\right)\left(\omega_{\alpha}+\eta_{\alpha}\Omega_0\right)dt-\eta_0
X_0 \Omega_0 dt-\eta_jX_j H^*\Omega_0
dt-\eta_{\alpha}\Omega_0X_{\alpha}H^* dt\\
-\left.\frac{\partial H^*}{\partial y^*_j} dy^*_j
\Omega_0+\Omega_0X_0H^* dt+\Omega_j X_j H^* dt +\Omega_{\alpha}
X_{\alpha} H^* dt+\frac{\partial H^*}{\partial y^*_j} \Delta y^*_j
dt\right]
\end{eqnarray*}
\end{small}
Taking into account (2.16-17) and (3.4), we obtain
\begin{small}
\begin{equation*}
\begin{split}
0&=\oint \left[\left\{\dot{y}^*_i-C^j_{0i}y^*_j-C^j_{ri}\eta_r
y^*_j-y^*_j(C_{0\alpha}^j+C_{r \alpha}^j \eta_r)\frac{\partial
\phi_{\alpha}}{\partial \eta_i}+X_i H^*+\frac{\partial
\phi_{\alpha}}{\partial \eta_i}X_{\alpha}H^*\right\}\Omega_i
dt\right.\\
&-\Delta y^*_j(\eta_j-\frac{\partial H^*}{\partial
y^*_j})dt+\left\{y^*_j(C_{0\alpha}^j+C_{r \alpha}^j
\eta_r)\frac{\partial \phi_{\alpha}}{\partial \eta_i}-X_i
H^*-\frac{\partial \phi_{\alpha}}{\partial
\eta_i}X_{\alpha}H^*\right\}\eta_i \Omega_0 dt\\
&\left.+\left\{-y^*_j(C_{0\alpha}^j+C_{r \alpha}^j
\eta_r)\eta_{\alpha} \Omega_0 dt-\eta_{\alpha}X_{\alpha} H^*
\Omega_0 dt-\frac{\partial H^*}{\partial y^*_j} dy^*_j
\Omega_0+\eta_{\alpha} \Omega_0 X_{\alpha} H^* dt\right\}\right]
\end{split}
\end{equation*}
\end{small}
which, by using (3.11-13), reduces to
\begin{equation}
\begin{split}
0&=\oint \left[\left\{\dot{y}^*-y^*_j(K^j_{0
i}+K^j_{ri}\eta_r)+X^*_i H^*\right\}\Omega_i dt-\Delta
y^*_j(\eta_j-\frac{\partial H^*}{\partial y^*_j})dt\right.\\
&+\left.\left\{y^*_j(C_{0\alpha}^j+C_{r \alpha}^j
\eta_r)\frac{\partial \phi_{\alpha}}{\partial \eta_i}\eta_i-X^*_i
H^*\eta_i -y^*_j(C_{0\alpha}^j+C_{r \alpha}^j
\eta_r)\eta_{\alpha}-\frac{\partial H^*}{\partial
y^*_j}\dot{y}^*_j\right\}\Omega_0 dt\right]
\end{split}
\end{equation}
As required by the present investigation, we adopt the second
viewpoint for which $\delta ^* f_{\alpha}=0$ for nonlinear
nonholonomic dynamical system and multiply (3.9) with
$(\frac{\partial L}{\partial \eta_\alpha})^*$ we get
\begin{equation}
(A_i^{\alpha})^* \left(\frac{\partial L}{\partial
\eta_{\alpha}}\right)^*\omega_i=0\quad (i=1,...,m; \alpha=m+1,...,n)
\end{equation}
Since the relations (2.17) hold for all the $\omega_i$'s, we
integrate (6.5) along the closed curve $C$ and incorporate the
result into (6.4), we obtain
\begin{equation*}
\begin{split}
0&=\oint \left[\left\{\dot{y}_i^*-y^*_j(K^j_{0
i}+K^j_{ri}\eta_r)+X^*_i H^*-(A_i^{\alpha})^* \left(\frac{\partial
L}{\partial \eta_{\alpha}}\right)^*\right\}\Omega_i dt\right.\\
&-\Delta y^*_j(\eta_j-\frac{\partial H^*}{\partial y^*_j})dt
+\left.\left\{y^*_j(C_{0\alpha}^j+C_{r \alpha}^j
\eta_r)\frac{\partial \phi_{\alpha}}{\partial \eta_i}\eta_i-X^*_i
H^*\eta_i \right.\right.\\
&+\eta_i \Omega_0 (A_i^{\alpha})^* \left(\frac{\partial L}{\partial
\eta_{\alpha}}\right)^*\left.\left.-y^*_j(C_{0\alpha}^j+C_{r
\alpha}^j \eta_r)\eta_{\alpha}-\frac{\partial H^*}{\partial
y^*_j}\dot{y}^*_j\right\}\Omega_0 dt\right]
\end{split}
\end{equation*}
which, in view of the equations (6.1) and (6.2), takes the form
\begin{equation*}
\begin{split}
0&=\oint \left[\left\{\phi_i-y^*_j(K^j_{0 i}+K^j_{ri}\eta_r)+X^*_i
H^*-(A_i^{\alpha})^* \left(\frac{\partial L}{\partial
\eta_{\alpha}}\right)^*\right\}\Omega_i \right.\\
&-\Delta y^*_j(\psi_j-\frac{\partial H^*}{\partial
y^*_j})+\left.\left\{y^*_j(C_{0\alpha}^j+C_{r \alpha}^j
\eta_r)\frac{\partial \phi_{\alpha}}{\partial \eta_i}\eta_i-X^*_i
H^*\eta_i \right.\right.\\
&+\eta_i \Omega_0 (A_i^{\alpha})^* \left(\frac{\partial L}{\partial
\eta_{\alpha}}\right)^*\left.\left.-y^*_j(C_{0\alpha}^j+C_{r
\alpha}^j \eta_r)\eta_{\alpha}-\frac{\partial H^*}{\partial
y^*_j}\dot{y}^*_j\right\}\Omega_0 \right]dt
\end{split}
\end{equation*}

In order that the equations (6.1) and (6.2) must be satisfied for
the asynchronous variation in which the quantities $\Omega_0,
\Omega_i$ and $\Delta y^*_j$ are arbitrary, the coefficient of each
of these must vanish.\\ This implies that
\begin{equation*}
\phi_i=-X_i^* H^*+(K^j_{0 i}+K^j_{ri}\eta_r)y_j^*+(A_i^{\alpha})^*
\left(\frac{\partial L}{\partial \eta_{\alpha}}\right)^*; \quad
\psi_i=\frac{\partial H^*}{\partial y^*_i}
\end{equation*}
\[
(j=1,2,...,m;\alpha=m+1,...,n;r=1,2,...,n)
\]
which yield the desired values of the functions $\phi_i$ and
$\psi_i$ and hence leads to the Poincar\'{e}-Hamilton equations of
motion. Also we have
\begin{eqnarray*}
y^*_j(C_{0\alpha}^j+C_{r \alpha}^j \eta_r)\frac{\partial
\phi_{\alpha}}{\partial \eta_i}\eta_i-X^*_i H^*\eta_i +\eta_i
\Omega_0 (A_i^{\alpha})^* \left(\frac{\partial L}{\partial
\eta_{\alpha}}\right)^*-y^*_j(C_{0\alpha}^j+C_{r \alpha}^j
\eta_r)\eta_{\alpha}\\
-\frac{\partial H^*}{\partial
y^*_j}\dot{y}^*_j=0
\end{eqnarray*}
Thus the preceding analysis can be summarized in the following:
\begin{theorem}
If the line integral (5.9) is invariant under a deformation of an
arbitrary curve $C$ along the tube of real trajectories, then the
motion of the conservative nonlinear nonholonomic dynamical system
is determined by the Poincar\'{e}-Hamilton(PH)  equations (4.17)
together with (4.7),provided that the relations (3.8) hold.
\end{theorem}

It is remarkable to note that Theorems 5.1 and 6.1 furnish the
necessary and sufficient condition which allow us to connect the
theory of integral invariants with the theory of
Poincar\'{e}-Hamiltonian systems. Thus, these integrals, as
discussed in
\cite{Naseer94,Naseerpreprint,Gantmacher,Pars,Whittaker,ott}, are
very important in analytical dynamics in the sense that they provide
another foundation for not only holonomic or linear nonholonomic
systems but also furnishes a foundation for nonlinear nonholonomic
dynamical systems whose motion is determined by the PH-system of
equations (4.17) with (4.7).

\section{A Generalization of The Poincar\'{e} Linear Integral Invariant}
We now turn to a generalization of a theorem analogous to the
theorem of Poincar\'{e} \cite{Naseerpreprint,Pars} as a special case
of the results established in the last section. To achieve this aim,
we proceed with the assumption that the variations under
consideration are synchronous. This implies that the quantity
$\Omega_0=\Delta t\equiv0$. From (2.17) and (2.21), it follows that
$\Omega_p=\omega_p$ and $\Delta \eta_p=\delta \eta_p$. Thus
$\Delta\equiv d$ holds and therefore $d \delta=\delta d$ holds.

Under such  conditions, we may restate Theorem 5.1 as follows
\begin{theorem}
The line integral
\begin{equation}
I_1=\oint _C y^*_j \omega_j
\end{equation}
along any closed curve $C$ that consists of the simultaneous states
of the system does not change with arbitrary deformation of this
curve along the tube of real trajectories of the conservative
nonlinear nonholonomic dynamical system which is described by the PH
system of equations (4.17) with (4.7).
\end{theorem}
The proof of the theorem adopts a procedure similar to that
discussed in the preceding sections but without requiring  Lemma
2.4. The invariance of the integral (7.1) can be obtained by
considering the simultaneous states of the dynamical system during
its motion in the "Reduced Phase-space" of $2m$ variables
$(x_j,y_j^*)$ and using the fact that, contrary to the relation
(2.23), the $d-$operation and integration commute with each other.
Provided we have  synchronous variation, we may assert the converse
of Theorem 3 in the following:
\begin{theorem}
If the line integral $I_1$, given by (7.1), remains invariant under
an arbitrary deformation along the tube of real trajectories of any
closed curve $C$ consisting of the simultaneous states of a
conservative nonlinear nonholonomic dynamical system, then the
motion of the system is determined by the PH system of equations
(4.17) with (4.7).
\end{theorem}

We remark that the results above show that the theory of integral
invariants forms another basis for both the Hamiltonian dynamics of
holonomic systems, linear nonholonomic systems and for nonlinear
nonholonomic dynamical systems as well.

In order to demonstrate the utility of the four theorems, we derive
special cases of our general results that are analogous to well
known results.

\begin{itemize}
\item[(i)] Suppose that all the $x$'s are the Lagrangian coordinates
and the $\eta$'s are the generalized velocities $\dot{x}$'s. In this
case, the relations (2.17) reduce to the result given in
\cite{Vujanovic} and \cite{Whittaker} by Vujanovic and Whittaker,
respectively; the operators $X_0$ and $X_p$'s becomes $\frac
{\partial}{\partial t}$ and $\frac {\partial}{\partial x_p}$'s.
Consequently all the $C_{0q}^p$'s and $C_{qp}^r$'s vanish. Theorem
7.1 and 7.2 furnish the results analogous to those obtained in
\cite{Cartan}, \cite{Gantmacher}, \cite{Pars}, and \cite{Whittaker}
by Cartan, Gantmacher, Pars and Whittaker, respectively while
Theorems 7.1 and 7.2 subsume the results that are discussed in
\cite{Poincare} and \cite{Poincare1901} by Poincar\'{e} and in
\cite{Pars} by Pars. It is to be noted that our results are
analogous to these but the content is quite different, since our
system is nonlinear nonholonomic.
\item[(ii)] If the group variables are the quasi-variables
(nonholonomic coordinates) $\pi$'s then the $\eta$'s becomes
$\dot{\pi}$'s and the relations (2.10) express these as
 non-integrable linear combinations of the quasi-velocities and all
the $C_{qp}^r$'s reduce to Hamel-Boltzmann's three indexed symbols
$\gamma_{pq}^r$. In this case our theorems subsume the results
analogous to those obtained in \cite{Djukic76} by Djukic.
\end{itemize}

%\section{APPENDIX: Asynchronous Variation of Functional (5.1)}
\renewcommand{\theequation}{A-\arabic{equation}}
% redefine the command that creates the equation no.
\setcounter{equation}{0}  % reset counter
\section*{APPENDIX\newline Asynchronous Variation of Functional (5.1)}  % use *-form to suppress numbering
Performing the $\Delta$-variation of (5.1) according to the Lemma
2.4, we have
\begin{equation*}
\begin{split}
\Delta S&=\int_{t_1}^{t_2}(\Delta L+L \dot{\Omega}_0)dt\\
&=\int_{t_1}^{t_2}(\frac{\partial L}{\partial \eta_p} \Delta
\eta_p+\Omega_\mu X_\mu L+(L \Omega_0\dot{)}-\dot{L}\Omega_0)dt
\end{split}
\end{equation*}
where we have used Def. 2.3 and the identity $(L\Omega_0\dot{)}=L
\dot{\Omega}_0+\dot{L}\Omega_0$. Taking into account (2.16), (2.17)
and (2.21), the variation $\Delta S$ of $S$ becomes
\begin{multline}
\Delta S=\int_{t_1}^{t_2}(\frac{\partial L}{\partial \eta_p} \delta
\eta_p+\omega_p X_p L)dt+\int_{t_1}^{t_2}(\frac{\partial L}{\partial
\eta_p}\dot{\eta}_p \Omega_0+\eta_\mu \Omega_0 X_\mu L-\Omega_0
\dot{L})dt\\
+\int_{t_1}^{t_2}(L \Omega_0\dot{)}dt
\end{multline}
Let
\begin{equation*}
I=\int_{t_1}^{t_2}(\frac{\partial L}{\partial \eta_p} \delta
\eta_p+\omega_p X_p L)dt,
\end{equation*}
separating the sum over the index $p$ from 1 to $n$ into the sums
over $j$ from 1 to $m$ and over $\alpha$ from $(m+1)$ to $n$ and
using the equation of motion (4.1) to get
\begin{equation*}
I=\int_{t_1}^{t_2}\left\{\left(\frac{\partial L^*}{\partial
\eta_j}-\frac{\partial L}{\partial \eta_\alpha}\frac{\partial
\phi_\alpha}{\partial \eta_j}\right) \delta \eta_j+\frac{\partial
L}{\partial \eta_\alpha}\delta \eta_\alpha+\omega_p X_p L\right\}dt.
\end{equation*}
Here the `$*$' over the quantities show that they are expressed in
terms of the independent parameters $\eta_j$ of real displacement.

Following the second viewpoint and using relation (2.20), we have
\begin{equation*}
\begin{split}
I&=\int_{t_1}^{t_2}\frac{\partial L^*}{\partial
\eta_j}(\dot{\omega}_j+C^j_{0q}\omega_q+C^j_{qr}\eta_q \omega_r)dt-
\int_{t_1}^{t_2}\frac{\partial L}{\partial
\eta_\alpha}\frac{\partial \phi_\alpha}{\partial
\eta_j}(\dot{\omega}_j+C^j_{0q}\omega_q\\
&+C^j_{qr}\eta_q \omega_r)dt+ \int_{t_1}^{t_2}\left\{\frac{\partial
L}{\partial
\eta_\alpha}(\dot{\omega}_\alpha+C^\alpha_{0q}\omega_q+C^\alpha_{qr}\eta_q
\omega_r)+\omega_p X_p L\right\}dt.
\end{split}
\end{equation*}
Integrating by parts, the first term of each integral on the right
hand side of the last result, we get
\begin{multline*}
I=\left.\left(\frac{\partial L^*}{\partial
\eta_j}\omega_j\right)\right|^{t_2}_{t_1}-\int_{t_1}^{t_2}\frac{d}{dt}(\frac{\partial
L^*}{\partial \eta_j})\omega_j dt +\int_{t_1}^{t_2}\frac{\partial
L^*}{\partial \eta_j}(C^j_{0q}\omega_q+C^j_{qr}\eta_q
\omega_r)dt\\
-\left.\left(\frac{\partial L}{\partial \eta_\alpha}\frac{\partial
\phi_\alpha}{\partial \eta_j}\omega_j\right)\right|^{t_2}_{t_1}
+\int_{t_1}^{t_2}\omega_j\frac{d}{dt}(\frac{\partial L}{\partial
\eta_\alpha}\frac{\partial \phi_\alpha}{\partial
\eta_j})dt-\int_{t_1}^{t_2}\frac{\partial L}{\partial
\eta_\alpha}\frac{\partial \phi_\alpha}{\partial
\eta_j}(C^j_{0q}\omega_q+C^j_{qr}\eta_q \omega_r)dt\\
+ \left.(\frac{\partial L}{\partial
\eta_\alpha}\omega_\alpha)\right|^{t_2}_{t_1}
-\int_{t_1}^{t_2}\frac{d}{dt}(\frac{\partial L}{\partial
\eta_\alpha})\omega_\alpha dt+\int_{t_1}^{t_2}\left(\frac{\partial
L}{\partial \eta_\alpha}(C^\alpha_{0q}\omega_q+C^\alpha_{qr}\eta_q
\omega_r)+\omega_p X_p L\right)dt
\end{multline*}
which by use of (3.4), reduces to
\begin{eqnarray*}
I&=&\left.\left(\frac{\partial L^*}{\partial
\eta_j}\omega_j\right)\right|^{t_2}_{t_1}-\int_{t_1}^{t_2}\left\{\frac{d}{dt}(\frac{\partial
L^*}{\partial \eta_j})\omega_j -\frac{\partial L^*}{\partial
\eta_j}(C^j_{0q}\omega_q+C^j_{qr}\eta_q \omega_r)\right.\\
&-&\omega_j \frac{\partial L}{\partial
\eta_\alpha}\frac{d}{dt}(\frac{\partial \phi_\alpha}{\partial
\eta_j})-\omega_j \frac{d}{dt}(\frac{\partial L}{\partial
\eta_\alpha})\frac{\partial \phi_\alpha}{\partial
\eta_j})+\frac{\partial L}{\partial \eta_\alpha}\frac{\partial
\phi_\alpha}{\partial \eta_j}(C^j_{0q}\omega_q+C^j_{qr}\eta_q
\omega_r)\\
&+&\left.\frac{d}{dt}(\frac{\partial L}{\partial
\eta_\alpha})\frac{\partial \phi_\alpha}{\partial
\eta_j}\omega_j-\frac{\partial L}{\partial
\eta_\alpha}(C^\alpha_{0q}\omega_q+C^\alpha_{qr}\eta_q
\omega_r)-\omega_p X_p L \right\}dt
\end{eqnarray*}

Again breaking the sum over the indices $q$ and $r$ from $1$ to $n$
into the sums over the indices $j,k$ from $1$ to $m$ and over the
indices $\alpha$, $\beta$ from $(m+1)$ to $n$, the last expression
takes the form
\begin{eqnarray*}
I&=&\left.\left(\frac{\partial L^*}{\partial
\eta_j}\omega_j\right)\right|^{t_2}_{t_1}-\int_{t_1}^{t_2}\left\{\frac{d}{dt}
(\frac{\partial
L^*}{\partial \eta_j})\omega_j -\frac{\partial L^*}{\partial
\eta_j}(C^j_{0k}\omega_k+C^j_{0\alpha}\omega_\alpha+C^j_{qk}\eta_q\omega_k\right.\\
&+& C^j_{q\alpha}\eta_q\omega_\alpha)-\omega_j \frac{\partial
L}{\partial \eta_\alpha}\frac{d}{dt}(\frac{\partial
\phi_\alpha}{\partial \eta_j})+\frac{\partial L}{\partial
\eta_\alpha}\frac{\partial \phi_\alpha}{\partial
\eta_j}(C^j_{0k}\omega_k+C^j_{0\beta}\omega_\beta+
C^j_{qk}\eta_q\omega_k\\
&+&C^j_{q\beta}\eta_q\omega_\beta)-\left.\frac{\partial L}{\partial
\eta_\alpha}(C^\alpha_{0j}\omega_j+C^\alpha_{0\beta}\omega_\beta+
C^\alpha_{qj}\eta_q\omega_j+C^\alpha_{q\beta}\eta_q\omega_\beta)-\omega_p
X_p L \right\}dt
\end{eqnarray*}
Interchanging the indices $j$ and $k$ in the second and fourth terms
of the integrand, using (3.4) and rearranging the terms, we find
that
\begin{small}
\[
\begin{split}
I&=\left.\left(\frac{\partial L^*}{\partial
\eta_j}\omega_j\right)\right|^{t_2}_{t_1}-\int_{t_1}^{t_2}\left[\frac{d}{dt}(\frac{\partial
L^*}{\partial \eta_j})\omega_j -\frac{\partial L^*}{\partial
\eta_k}\left\{(C^k_{0j}+C^k_{0\alpha}\frac{\partial
\phi_\alpha}{\partial \eta_j}) +\eta_q(C^k_{qj}
+C^k_{q\alpha}\frac{\partial \phi_\alpha}{\partial \eta_j})\right\}
\omega_j\right.\\
 &-\omega_j \frac{\partial L}{\partial
\eta_\alpha}\frac{d}{dt}(\frac{\partial \phi_\alpha}{\partial
\eta_j}) +\frac{\partial L}{\partial \eta_\alpha}\frac{\partial
\phi_\alpha}{\partial
\eta_k}\left\{(C^k_{0j}+C^k_{0\beta}\frac{\partial \phi_\beta}
{\partial \eta_j})+(C^k_{qj}+C^k_{q\beta}\frac{\partial
\phi_\beta}{\partial \eta_j}) \eta_q\right\}\omega_j
\\
&-\left.\frac{\partial L}{\partial
\eta_\alpha}\left\{(C^\alpha_{0j}+C^\alpha_{0\beta}\frac{\partial
\phi_\beta}{\partial
\eta_j})+(C^\alpha_{qj}+C^\alpha_{q\beta}\frac{\partial
\phi_\beta}{\partial \eta_j})\eta_q\right\}\omega_j-\omega_p X_p L
\right]dt
\end{split}
\]
\end{small}
Taking into account equations (3.11) and (3.12), the last result
reduces to
\begin{eqnarray}
I&=& \left. \frac{\partial L^*}{\partial
\eta_j}\omega_j\right|^{t_2}_{t_1}-\int_{t_1}^{t_2}\left[\frac{d}{dt}(\frac{\partial
L^*}{\partial \eta_j})\omega_j -\frac{\partial L^*}{\partial
\eta_k}(K_{0j}^{k}+K_{qj}^{k}\eta_q)\omega_j\right.\\
&-& \left.\omega_j(\frac{\partial L}{\partial
\eta_\alpha})^*\left\{\frac{d}{dt}(\frac{\partial
\phi_\alpha}{\partial \eta_j})-\frac{\partial \phi_\alpha}{\partial
\eta_k}(K_{0j}^{k}+K_{qj}^{k}\eta_q)+(K_{0j}^{\alpha}+K_{qj}^{\alpha}\eta_q)\right\}\omega_j\right.\notag\\
&-&\omega_p X_pL\big]dt\notag
\end{eqnarray}
where the "$*$" over the quantities $(\frac{\partial L}{\partial
\eta_\alpha})^*$ indicates that they are expressed in terms of the
independent parameters $\eta_j$.

Let us consider the term $\omega_pX_pL$ which, in view of (4.4), can
be written as
\begin{eqnarray*}
\omega_p X_p L&=& \omega_p X_p L^*-\omega_p \frac{\partial
L}{\partial
\eta_\alpha} X_p \phi_\alpha\\
&=& \omega_j X_j L^*+\omega_\alpha X_\alpha
L^*-\omega_j\frac{\partial L}{\partial \eta_\alpha}X_j
\phi_\alpha-\omega_\beta\frac{\partial L}{\partial
\eta_\alpha}X_\beta \phi_\alpha\\
&=&\omega_j X_j L^*+\frac{\partial \phi_\alpha}{\partial
\eta_j}\omega_j X_\alpha L^*-\omega_j\frac{\partial L}{\partial
\eta_\alpha}X_j\phi_\alpha-\frac{\partial \phi_\beta}{\partial
\eta_j}\omega_j\frac{\partial L}{\partial \eta_\alpha}X_\beta
\phi_\alpha
\end{eqnarray*}
where to obtain this result we have separated the sum over the index
$p=1,2,...,n$ into the sums over $j=1,2,...,m$ and
$\alpha=m+1,...,n$ and also used the relation (4.3). Simplifying the
last expression, we get
\begin{equation*}
\omega_p X_p L=\omega_j(X_j +\frac{\partial \phi_\alpha}{\partial
\eta_j}X_\alpha)L^*-\omega_j \frac{\partial L}{\partial
\eta_\alpha}(X_j+\frac{\partial \phi_\beta}{\partial
\eta_j}X_\beta)\phi_\alpha,
\end{equation*}
which together with (3.13), becomes
\begin{equation}
\omega_p X_p L=\omega_j X_j^*L^*-\omega_j(\frac{\partial L}{\partial
\eta_\alpha})^* X^*_j \phi_\alpha,
\end{equation}
where we have expressed all the quantities in terms of the
independent parameters $\eta_j$'s of real displacement. This allows
us to write (A.1) as
\begin{small}
\begin{eqnarray*}
I&=& \left. \left(\frac{\partial L^*}{\partial
\eta_j}\omega_j\right)\right|^{t_2}_{t_1}-\int_{t_1}^{t_2}\left[\frac{d}{dt}(\frac{\partial
L^*}{\partial \eta_j}) -\frac{\partial L^*}{\partial
\eta_k}(K_{0j}^{k}+K_{qj}^{k}\eta_q)-X^*_j L^*\right.\\
&-&\left.(\frac{\partial L}{\partial
\eta_\alpha})^*\left\{\frac{d}{dt}(\frac{\partial
\phi_\alpha}{\partial \eta_j})-\frac{\partial \phi_\alpha}{\partial
\eta_k}(K_{0j}^{k}+K_{qj}^{k}\eta_q)+(K_{0j}^{\alpha}+K_{qj}^{\alpha}\eta_q)-X^*_j\phi_\alpha\right\}\right]\omega_j
dt
\end{eqnarray*}
\end{small}
\begin{equation}
 I=\left. \left(\frac{\partial L^*}{\partial
\eta_j}\omega_j\right)\right|^{t_2}_{t_1}-\int_{t_1}^{t_2}\left[\frac{d}{dt}(\frac{\partial
L^*}{\partial \eta_j}) -(K_{0j}^{k}+K_{qj}^{k}\eta_q)\frac{\partial
L^*}{\partial \eta_k}-X^*_j L^*-(A^{\alpha}_j)^*(\frac{\partial
L}{\partial \eta_\alpha})^*\right]\omega_j dt
\end{equation}
which expresses the asynchronous variation of the action integral
(5.1).
\bibliographystyle{plain}
\bibliography{mechanics}

\begin{thebibliography}{10}

\bibitem{NaseerThesis}
Naseer Ahmed.
\newblock Some problems in the dynamics of nonholonomic systems.
\newblock {\em PhD Dissertation, Quaid-i-Azam University, Islamabad,
  Pakistan.}, 1986.

\bibitem{Naseer94}
Naseer Ahmed.
\newblock Integral invariants of a holonomic dynamical system.
\newblock {\em Appl. Math. Mech.}, 15(8):755--765, 1994.

\bibitem{Naseerpreprint}
Naseer Ahmed and Naeem ul~Haq.
\newblock Integral invariants of a linear nonholonomic dynamical system.
\newblock {\em Preprints, Quaid-i-Azam University, Islamabad, Pakistan.}

\bibitem{Arnold}
V.~I. Arnold.
\newblock {\em Mathematical methods of classical mechanics}.
\newblock Springer-Verlag, New York, 1978.
\newblock Translated from the Russian by K. Vogtmann and A. Weinstein, Graduate
  Texts in Mathematics, 60.

\bibitem{Gomes79}
F.~Benevent and J.~Gomes.
\newblock {P}oincar\'e-{C}artan integral invariant for constrained system.
\newblock {\em Ann. Phys.}, 118(2):467--489, 1979.

\bibitem{Blacall41}
C.~J. Blacall.
\newblock On volume integral invariants of nonholonomic dynamical systems.
\newblock {\em Amer. J. Math.}, 63:155--168, 1941.

\bibitem{Cartan}
E.~J. Cartan.
\newblock Leson sur le invariant int\'egraux.
\newblock {\em Hanmann, Paris}, 1921.

\bibitem{Chetaev41}
N.~G Chetaev.
\newblock On the equation of {P}oincar\'e.
\newblock {\em J. Appl. Math. Mech}, 5:253--262, 1941.

\bibitem{Djukic76}
Dj.~S. Djukic.
\newblock A variational principle involving a conditional extremum for the
  {H}amel-{B}oltzman equations of motion.
\newblock {\em Acta Mechanica}, 25:105--110, 1976.

\bibitem{Dobronravov}
V.~V. Dobronravov.
\newblock Integral invariants of the analytical mechanics in nonholonomic
  coordinates.
\newblock T. XLVI(5):196--199, 1945.
\newblock (in Russian).

\bibitem{Donder}
T.~Donder.
\newblock Sur le invariants int\'egraux felatifs et leur application a la
  physique mathmatique.
\newblock {\em Bulletin de l'Acad. Royale de Belgique}, pages 50--70, 1911.
\newblock (classe de sciences).

\bibitem{Gantmacher}
F.~Gantmacher.
\newblock {\em Lectures in Analytical Mechanics}.
\newblock Mir Publisher, Moscow, 1970.

\bibitem{GN-HP-NH-94}
Q.~K. Ghori and N.~Ahmed.
\newblock Hamilton's principle for nonholonomic systems.
\newblock {\em Z. Angew. Math. Mech.}, 74(2):137--140, 1994.

\bibitem{Lee47}
H.~C. Lee.
\newblock The universal integral invariants of {H}amiltonian systems and
  application to the theory of canonical transformations.
\newblock {\em Proc. Roy. Soc. Edin.}, 62(Part A):237--246, 1947.

\bibitem{Feng91}
Feng~Xiang Mei.
\newblock First integral and integral-invariant for nonholonomic systems.
\newblock {\em Chinese Sci. Bull.}, 36(24):2038--2042, 1991.

\bibitem{Neimark-Fufaev}
J.~I. Neimark and N.~A Fufaev.
\newblock {\em Dynamics of Nonholonomic Systems}.
\newblock Nauka, Moscow, 1967.
\newblock in Russian.

\bibitem{ott}
Edward Ott.
\newblock {\em Chaos in dynamical systems}.
\newblock Cambridge University Press, Cambridge, second edition, 2002.

\bibitem{Pars}
L.~A. Pars.
\newblock {\em A treatise on analytical dynamics}.
\newblock John Wiley \& Sons Inc., New York, 1965.

\bibitem{Poincare}
H.~Poincar{\'e}.
\newblock Les {M}\'ethodes {N}ouvelles de la {M}\'ecanique {C}\'eleste. {T}om
  {III},{D}over {P}ub. {I}nc., {N}.{Y}.
\newblock 1892.

\bibitem{Poincare1901}
H.~Poincar\'e.
\newblock Sur une forme nouvelle des \'equations de la m\'ecanique.
\newblock {\em C. R. Acad. Sc., Paris}, 132:369--371, 1901.

\bibitem{Savchenko}
A.Y. Savchenko and B.~Y. Zeldovich.
\newblock Speckle beams with nonzero vorticity and {P}oincar\'e-{C}artan
  invariant.
\newblock {\em J. Opt. Soc. Am. A}, 16:1665--1671, 1999.

\bibitem{Shao93}
L.~Shao-Kai.
\newblock Integral theory for the dynamics of nonlinear nonholonomic systems in
  non-inertial frame of reference.
\newblock {\em Appl. Math. Mech.}, 14(10):907--918, 1993.

\bibitem{Shu}
H.~B. Shu.
\newblock Application of the integral invariant of {P}oincar\'e-{C}artan to
  contact system.
\newblock {\em Computers math. Applic.}, 26(2):51--59, 1993.

\bibitem{Ting}
L.~U. Ting.
\newblock On the application of the integral invariants and decay laws of
  vorticity distributions.
\newblock {\em J. Fluid Mech.}, 127:497--506, 1993.

\bibitem{Vujanovic}
B.~Vujanovic.
\newblock Conservation laws of dynamical systems via d'{A}lembert's principle.
\newblock {\em Int. J. Nonlinear Mech.}, 13:185--197, 1978.

\bibitem{Whittaker}
T.~Whittaker, E.
\newblock Treatise on the analytical dynamics of particles and rigid bodies.
\newblock {\em Cambridge Univ. Press, Cambridge}, 1904.

\bibitem{Yarosheuk}
V.~A. Yarosheuk.
\newblock New cases of existence of an integral invariant.
\newblock {\em Univ. Ser. I. Mat. Mekh., Vestnik Moskov}, 6:26--30, 1992.
\newblock (in Russian).

\end{thebibliography}
\end{document}